\documentclass[preprint,prd,nofootinbib,tightenlines,amsmath]{revtex4}

\usepackage{axodraw} 
\usepackage{bm} 
\usepackage{graphicx}

\begin{document}

\baselineskip=15pt
\parskip=5pt

\hfill SMU-HEP-04-06

\vspace*{10ex}

\title{Probing  $\bm{CP}$  Violation in
$\bm{\Omega\to\Lambda K\to p\pi K}$  Decay}

\author{Jusak Tandean}
\email{jtandean@mail.physics.smu.edu}

\affiliation{
Department of Physics, Southern Methodist University,
Dallas, Texas 75275-0175
\vspace{5ex} \\}


\begin{abstract}
The sum of the $CP$-violating asymmetries  $A_\Omega^{}$  and
$A_\Lambda^{}$  in the decay sequence
$\Omega\to\Lambda K,\,$ $\Lambda\to p\pi$
is presently being measured by the E871 experiment.
We evaluate contributions to  $A_\Omega^{}$  from the standard model
and from possible new physics, and find them to be smaller
than the corresponding contributions to~$A_\Lambda^{}$, although not
negligibly so.
We also show that the partial-rate asymmetry in  $\Omega\to\Lambda K$
is nonvanishing due to final-state interactions.
Taking into account constraints from kaon data, we discuss
how the upcoming result of E871 and future measurements may probe
the various contributions to the observables.
\end{abstract}


\maketitle

\section{Introduction\label{intro}}

The question of the origin of  $CP$  violation remains one of
the outstanding puzzles in particle physics.
Although  $CP$ violation has now been seen in a number of processes in
the kaon and $B$-meson systems~\cite{cpx}, it is still far from clear
whether its explanation lies exclusively within the picture provided
by the standard model~\cite{buras1}.
To pin down the sources of $CP$ violation, it is essential to observe
it in many other processes.

Hyperon nonleptonic decays provide an environment where it is possible
to make additional observations of $CP$ violation~\cite{hcpvt,hypercp}.
Currently, there are $CP$-violation searches in such processes  
being conducted by the HyperCP~(E871) Collaboration at  Fermilab.
Its main reactions of interest are the decay chain   
$\,\Xi^-\to\Lambda\pi^-,\,$  $\,\Lambda\to p\pi^-\,$  
and its antiparticle counterpart~\cite{hypercp}.  
A different, but related, system also being studied by HyperCP
involves the spin-$\frac{3}{2}$  hyperon  $\Omega^-$, namely
the sequence
$\,\Omega^-\to\Lambda K^-,\,$  $\,\Lambda\to p\pi^-\,$  and its
antiparticle process~\cite{lu}.
For each of these decays, the decay distribution in the rest frame of
the parent hyperon with known polarization  $\bm{w}$ has the form
\begin{equation}
\frac{{\rm d}\Gamma}{{\rm d}\Omega}  \,\sim\,
1 +\alpha\, \bm{w}\cdot\hat{\bm{p}}   \,\,,
\end{equation}
where  ${\rm d}\Omega$  is the final-state solid angle,  $\hat{\bm{p}}$
is the unit vector of the daughter-baryon momentum,  and  $\alpha$  is
the parameter relevant to the $CP$ violation of interest.
In the case of  $\,\Omega\to\Lambda K\to p\pi K,\,$  the HyperCP
experiment is sensitive to the {\sl sum} of $CP$ violation in
the  $\Omega$  decay and  $CP$ violation in the  $\Lambda$  decay,
measuring~\cite{lu}
\begin{equation}   \label{A_OL}
A_{\Omega\Lambda}^{}  \,=\,
\frac{\alpha_\Omega^{} \alpha_\Lambda^{}
      - \alpha_{\overline{\Omega}} \alpha_{\overline{\Lambda}}}
     {\alpha_\Omega^{} \alpha_\Lambda^{}
      + \alpha_{\overline{\Omega}} \alpha_{\overline{\Lambda}}}
\,\simeq\,
A_\Omega^{} + A_\Lambda^{}   \,\,,
\end{equation}
where
\begin{equation}
A_\Omega^{}  \,\equiv\,
\frac{\alpha_\Omega^{}+\alpha_{\overline{\Omega}}}
     {\alpha_\Omega^{}-\alpha_{\overline{\Omega}}}   \,\,,
\hspace{3em}
A_\Lambda^{}  \,\equiv\,
\frac{\alpha_{\Lambda}^{}+\alpha_{\overline{\Lambda}}}
     {\alpha_{\Lambda}^{}-\alpha_{\overline{\Lambda}}}
\end{equation}
are the $CP$-violating asymmetries in  $\,\Omega\to\Lambda K\,$
and  $\,\Lambda\to p\pi,\,$  respectively.
Similarly, the observable it measures in
$\,\Xi\to\Lambda\pi\to p\pi\pi\,$  is
$\,A_{\Xi\Lambda}^{}\simeq A_\Lambda^{} + A_\Xi^{}\,$~\cite{hypercp}.

On the theoretical side, $CP$ violation in  $\,\Lambda\to p\pi\,$  and
$\,\Xi\to\Lambda\pi\,$  has been extensively
studied~\cite{hcpvt,dhp,hcpvt',NP,tv4,t6}.
In contrast, the literature on  $CP$  violation in  $\Omega$  decays
is minimal, perhaps the only study being Ref.~\cite{tv2} which deals 
with the partial-rate asymmetry in  $\,\Omega\to\Xi\pi.\,$
There is presently no data available or experiment being done on
this rate asymmetry.
In view of the upcoming measurement of  $A_{\Omega\Lambda}^{}$  by
HyperCP, it is important to have theoretical expectations of this
observable.
Clearly, the information to be gained from  $A_{\Omega\Lambda}^{}$
will complement that from  $A_{\Xi\Lambda}^{}$.
Since the estimates of  $A_\Lambda^{}$  and  $A_\Xi^{}$  within and
beyond the standard model (SM) have been updated very recently in
Refs.~\cite{tv4,t6}, in this paper we focus on  $A_\Omega^{}$.

We begin in Sec.~\ref{observables} by relating the observables of
interest in  $\,\Omega\to\Lambda K\,$  to the strong and $CP$-violating
weak phases in the decay amplitudes.
We discuss the role played by final-state interactions in this decay,
which not only affect  $A_\Omega^{}$,  but also cause its partial-rate
asymmetry to be nonvanishing, thereby providing another $CP$-violating
observable.
In Sec.~\ref{strong_phases}, we employ heavy-baryon chiral
perturbation theory ($\chi$PT) to calculate  $P$- and $D$-wave
amplitudes for baryon-meson scattering in channels with isospin
$\,I=\frac{1}{2}\,$   and strangeness  $\,S=-2.\,$
We use the derived amplitudes in a coupled-channel $K$-matrix
formalism to determine the strong parameters needed in evaluating
the $CP$-violating asymmetries.
In Sec.~\ref{A_sm}, we estimate  the asymmetries within the standard model.
Working in the framework of $\chi$PT, we calculate the weak phases
by considering factorizable and nonfactorizable contributions to
the matrix elements of the leading penguin operator.
Subsequently, we compare the resulting  $A_\Omega^{}$  with
$A_\Lambda^{}$, which was previously evaluated, as both asymmetries
appear in $A_{\Omega\Lambda}^{}$.
In Sec.~\ref{A_np}, we address contributions to the $CP$-violating
asymmetries  from  possible new physics, taking into account
constraints from $CP$  violation in the kaon system.
Specifically, we consider contributions induced by
chromomagnetic-penguin operators, which in certain models can be
enhanced compared to the SM effects.
Sec.~\ref{conclusion} contains our conclusions.

\section{Observables and phases\label{observables}}

The amplitudes for  $\,\Omega^-\to\Lambda K^-\,$  and
$\,\bar{\Omega}{}^+\to\bar{\Lambda}K^+\,$  each contain
parity-conserving  $P$-wave  and parity-violating  $D$-wave components,
with the former being empirically known to be dominant~\cite{pdb}.
They are related to the parameters  $\alpha_\Omega^{}$  and
$\alpha_{\overline{\Omega}}$  by
\begin{eqnarray}   \label{alpha}
\alpha_\Omega^{}  \,=\,
\frac{2\,{\rm Re}\bigl(p^*d\bigr)}{|p|^2+|d|^2}   \,\,,
\hspace{3em}
\alpha_{\overline{\Omega}}  \,=\,
\frac{2\,{\rm Re}\bigl(\bar{p}{}^*\bar{d}\bigr)}
{\bigl|\bar{p}\bigr|^2+\bigl|\bar{d}\bigr|^2}   \,\,,
\end{eqnarray}
where  $p$ and $d$  $\bigl(\bar{p}$ and $\bar{d}\bigr)$  are the $P$-
and $D$-wave components, respectively, for the  $\Omega^-$
$\bigl(\bar{\Omega}{}^+\bigr)$  decay.
Since both  $\Omega$  and  $\Lambda$  have  $\,I=0,\,$  each of
these decays is an exclusively  $\,|\Delta I|=\frac{1}{2}\,$  transition.

Before writing down the amplitudes in terms their phases, we note that
the strong phases in  $\,\Omega\to\Lambda\bar{K}\,$  are not generated
by the strong rescattering of  $\Lambda\bar{K}$  alone.
Watson's theorem for elastic unitarity~\cite{watson} does not apply
here, though it does in the cases of  $\,\Lambda\to p\pi\,$  and
$\,\Xi\to\Lambda\pi.\,$
Final-state interactions also allow
$\,\Omega\to\Xi\pi\to\Lambda\bar{K}\,$  to contribute,
yielding additional strong phases as well as weak ones, because
the channel  $\,\Xi\pi\leftrightarrow\Lambda\bar{K}\,$  is open at
the scattering energy  $\,\sqrt{s}=m_\Omega.\,$
Since the  $\,\Omega\to\Xi\pi,\Lambda\bar{K}\,$  rates
overwhelmingly dominate the $\Omega$ width~\cite{pdb}, we expect
other contributions via final-state rescattering to be negligible.

The requirements of  $CPT$  invariance and unitarity provide us
with a relationship between the amplitudes for  $\,\Omega\to B\phi\,$
and  its antiparticle counterpart.   
Thus, with  ${\cal M}_{\Omega\to B\phi}^{(L)}$  denoting the amplitude 
corresponding to  $B\phi$ being in a state with orbital angular 
momentum $L$,  we have
\begin{eqnarray}   \label{M_Bphi}   
(-1)^{L+1}\, {\cal M}_{\overline{\Omega}\to\overline{\Lambda}K}^{(L)}  \,=\,
{\cal S}_{\Lambda\Lambda}^{(L)}\, {\cal M}_{\Omega\to\Lambda\bar{K}}^{(L)*}
+ {\cal S}_{\Lambda\Xi}^{(L)}\, {\cal M}_{\Omega\to\Xi\pi}^{(L)*}   \,\,,
\end{eqnarray}
where  ${\cal S}_{BB'}^{(L)}$  is the element of the strong $S$-matrix
associated with the $L$ partial-wave of $\,B\phi\to B'\phi',\,$  and  
only the  $\,I=\frac{1}{2}\,$ component of the $\Xi\pi$ state is involved 
in the second term.
Assuming that  the  $\Xi\pi$  and  $\Lambda\bar{K}$  channels are the only
ones open, we can express the $S$-matrix  as~\cite{pilkuhn}
\begin{eqnarray}   \label{S}
{\cal S}  \,=\,
\left( \begin{array}{cc}   \displaystyle
{\cal S}_{\Xi\Xi}^{}  &  {\cal S}_{\Xi\Lambda}^{}
\vspace{1ex} \\   \displaystyle
{\cal S}_{\Lambda\Xi}^{}  &  {\cal S}_{\Lambda\Lambda}^{}
\end{array} \right)
\,=\,
\left( \begin{array}{cc}   \displaystyle
\hat{\eta}\, {\rm e}^{2{\rm i}\delta_{\Xi\pi}}  &
{\rm i}\sqrt{1-\hat{\eta}^2}\;
{\rm e}^{{\rm i}(\delta_{\Xi\pi}+\delta_{\Lambda K})}
\vspace{1ex} \\   \displaystyle
{\rm i}\sqrt{1-\hat{\eta}^2}\;
{\rm e}^{{\rm i}(\delta_{\Xi\pi}+\delta_{\Lambda K})}  &
\hat{\eta}\, {\rm e}^{2{\rm i}\delta_{\Lambda K}}
\end{array} \right)   \,\,,
\end{eqnarray}
where  $\hat{\eta}$  is the inelasticity factor and
$\delta_{B\phi}^{}$  denotes the phase shift in  $\,B\phi\to B\phi.\,$
Clearly  $\cal S$~is unitary,  and  each partial-wave has its own
$\cal S$.
Now, since  $\hat{\eta}$  is expected to be close to and smaller than~1,
it is convenient to introduce a parameter  $\varepsilon$  defined by  
\begin{eqnarray}
\hat{\eta}  \,=\,  1-2\varepsilon   \,\,,  
\end{eqnarray}
and so  $\varepsilon$  is positive and small.
Consequently, for  $\,L=1$ and $2$,\,  to first order in  
$\sqrt{\varepsilon}$  we have~\cite{wolfenstein'}
\begin{eqnarray}   \label{pd}
\begin{array}{c}   \displaystyle
p  \,=\,
{\rm e}^{{\rm i}\delta_{\Lambda K}^P}\, \Bigl(
p_\Lambda^{}\, {\rm e}^{{\rm i}\phi_\Lambda^P}
+ {\rm i}\sqrt{\varepsilon_P^{}}\,\, p_\Xi^{}\, {\rm e}^{{\rm i}\phi_\Xi^P}
\Bigr)   \,\,,
\hspace{3em}
d  \,=\,
{\rm e}^{{\rm i}\delta_{\Lambda K}^D}\, \Bigl(
d_\Lambda^{}\, {\rm e}^{{\rm i}\phi_\Lambda^D}
+ {\rm i}\sqrt{\varepsilon_D^{}}\,\, d_\Xi^{}\, {\rm e}^{{\rm i}\phi_\Xi^D}
\Bigr)   \,\,,
\vspace{2ex} \\   \displaystyle
\bar{p}  \,=\,
{\rm e}^{{\rm i}\delta_{\Lambda K}^P}\, \Bigl(
p_\Lambda^{}\, {\rm e}^{-{\rm i}\phi_\Lambda^P}
+ {\rm i}\sqrt{\varepsilon_P^{}}\,\, p_\Xi^{}\, {\rm e}^{-{\rm i}\phi_\Xi^P}
\Bigr)   \,\,,
\hspace{3em}
\bar{d}  \,=\,
-{\rm e}^{{\rm i}\delta_{\Lambda K}^D} \Bigl(
d_\Lambda^{}\, {\rm e}^{-{\rm i}\phi_\Lambda^D}
+ {\rm i}\sqrt{\varepsilon_D^{}}\,\, d_\Xi^{}\, {\rm e}^{-{\rm i}\phi_\Xi^D}
\Bigr)   \,\,,
\end{array}
\end{eqnarray}
where  $p_B^{}$  and  $d_B^{}$  are real, associated with
$\,\Omega\to B\phi,\,$  and  $\phi_B^{P,D}$  denote
the corresponding weak phases in the  $\,|\Delta I|=\frac{1}{2}\,$
amplitudes.

Putting together the results above, and keeping only the terms at
lowest order in small quantities, we obtain
\begin{equation}   \label{AO}
A_\Omega^{}  \,=\,
- \tan \bigl( \delta_{\Lambda K}^P-\delta_{\Lambda K}^D \bigr) \,
\sin \bigl( \phi_\Lambda^P-\phi_\Lambda^D \bigr)
- \frac{p_\Xi^{}}{p_\Lambda^{}}\, \sqrt{\varepsilon_P^{}}\,
 \sin \bigl( 2\phi_\Lambda^P-\phi_\Xi^P-\phi_\Lambda^D \bigr)
+ \frac{d_\Xi^{}}{d_\Lambda^{}}\, \sqrt{\varepsilon_D^{}}\,
 \sin \bigl( \phi_\Lambda^P-\phi_\Xi^D \bigr)   \,\,,
\end{equation}
where we have made use of the expectation that
$\delta_{\Lambda K}^{P,D}$,  $\phi_{\Lambda,\Xi}^{P,D}$,  and
$d_B^{}/p_B^{}$  are also small.
Unlike the strong phases in  $\Lambda$  and  $\Xi$  decays,
there are no data currently available for  $\delta_{\Lambda K}^{}$,
and so we will calculate them here.
To estimate the weak phases  $\phi_{\Lambda,\Xi}^{}$,  we will consider
contributions coming from the SM as well as from possible new physics.
As for  $p_B^{}$  and  $d_B^{}$, we will extract their approximate
values from data shortly, under the assumption of no final-state
interactions and no $CP$ violation.

Now, the presence of the $\sqrt{\varepsilon}$ terms with additional
weak and strong phases in the decay amplitudes in Eq.~(\ref{pd})
implies that the rate of  $\,\Omega\to\Lambda\bar{K},\,$
\begin{eqnarray}   \label{width}
\Gamma_{\Omega\to\Lambda\bar{K}}^{}  \,=\,
\frac{ \bigl|\bm{k}_\Lambda^{}\bigr| \,
      \bigl(E_\Lambda^{}+m_\Lambda^{}\bigr) }{12\pi\, m_\Omega^{}}
\bigl( |p|^2 + |d|^2 \bigr)   \,\,,
\end{eqnarray}
evaluated in the rest frame of $\Omega$,  is no longer identical to
that of  $\,\bar{\Omega}\to\bar{\Lambda}K.\,$
Hence these decays yield another $CP$-violating observable, namely
the partial-rate asymmetry
\begin{eqnarray}   \label{deltao}
\Delta_\Omega^{}  \,=\,
\frac{ \Gamma_{\Omega\to\Lambda\bar{K}}^{}
      - \Gamma_{\overline{\Omega}\to\overline{\Lambda}K}^{} }
     { \Gamma_{\Omega\to\Lambda\bar{K}}^{}
      + \Gamma_{\overline{\Omega}\to\overline{\Lambda}K}^{} }   \,\,.
\end{eqnarray}
It follows that  to leading order
\begin{eqnarray}   \label{DeltaO}
\Delta_\Omega^{}  \,=\,
\frac{2\,p_\Xi^{}}{p_\Lambda^{}}\, \sqrt{\varepsilon_P^{}}\,\,
\sin \bigl( \phi_\Lambda^P-\phi_\Xi^P \bigr)   \,\,.
\end{eqnarray}
We will also estimate this asymmetry below.\footnote{In Ref.~\cite{tv2}
the partial-rate asymmetry in  $\,\Omega\to\Xi\pi\,$  was evaluated
under the assumption that  $\,\varepsilon=0.\,$
}
Since   $\Delta_\Omega^{}$  results from the interference of $P$-wave
amplitudes, a~future measurement of it  will probe $CP$
violation in the underlying parity-conserving interactions.
We note that the strong parameters entering  Eq.~(\ref{DeltaO}),
and the second and third terms in Eq.~(\ref{AO}), are not the strong
phases, but  $\varepsilon_{P,D}^{}$.

Before ending this section, we determine the values of  $p_B^{}$ and
$d_B^{}$  which are needed in Eqs.~(\ref{AO}) and~(\ref{DeltaO}),
and also in evaluating the weak phases.
To do so, we apply the measured values of $\alpha$ and $\Gamma$,
as well as of the masses involved, in the corresponding formulas,
as those in Eqs.~(\ref{alpha}) and~(\ref{width}), assuming that
the strong and weak phases are zero.
The experimental values of  $\Gamma$  for
$\,\Omega\to\Lambda\bar{K},\Xi\pi\,$  are well determined, but those
of  $\alpha$ are not~\cite{pdb}.
HyperCP is currently also measuring $\alpha_\Omega^{}$,
in  $\,\Omega\to\Lambda\bar{K},\,$  with much better precision, and
has reported~\cite{lu} preliminary results of
$\,\alpha_\Omega^{}=(1.84\pm0.46\pm0.04)\times10^{-2}\,$  and
$\,\alpha_\Omega^{}=(2.01\pm0.17\pm0.04)\times10^{-2}.\,$
Applying the PDG averaging procedure~\cite{pdb} to all
the experimental results, including the preliminary ones from
HyperCP, yields the average  $\,\alpha_\Omega^{}=0.020\pm0.002,\,$
which we adopt in the following.
In the case of  $\,\Omega\to\Xi\pi,\,$  we use the data given by
the PDG~\cite{pdb}, and also
\begin{eqnarray}   \label{|Xp>}
|\Xi\pi\rangle  \,=\,
\sqrt{\mbox{$\frac{2}{3}$}}\, \bigl| \Xi^0\pi^- \bigr\rangle
+ \mbox{$\frac{1}{\sqrt3}$}\, \bigl| \Xi^-\pi^0 \bigr\rangle
\end{eqnarray}
to project out the  $\,|\Delta I|=\frac{1}{2}\,$  amplitudes.
Thus we extract
\begin{eqnarray}   \label{px,dx}
\begin{array}{c}   \displaystyle
p_\Lambda^{}  \,=\,  3.73 \pm 0.03   \,\,,  \hspace{3em}
d_\Lambda^{}  \,=\,  0.037 \pm 0.004   \,\,,
\vspace{1ex} \\   \displaystyle
p_\Xi^{}  \,=\,  2.00 \pm 0.03   \,\,,  \hspace{3em}
d_\Xi^{}  \,=\,  0.08 \pm 0.12   \,\,,
\end{array}
\end{eqnarray}
all in units of  $G_{\rm F}^{}m_{\pi^+}^2$, with  $G_{\rm F}^{}$
being the Fermi coupling constant.

\section{Strong phases and inelasticity factors\label{strong_phases}}
  
To calculate the strong parameters needed in Eq.~(\ref{AO}), we take
a $K$-matrix approach~\cite{pilkuhn}.
Furthermore, we include the contributions of other  $B\phi$  states
with  $\,I=\frac{1}{2}\,$  and  $\,S=-2,\,$  namely  $\Sigma\bar{K}$
and  $\Xi\eta$,  which are coupled to  $\Lambda\bar{K}$  and  $\Xi\pi$
through unitarity constraints.
Although at  $\,\sqrt{s}=m_\Omega^{}\,$  the  $\Sigma\bar{K}$  and
$\Xi\eta$  channels are below their thresholds,  it is important to
incorporate their contributions to the open ones.
Such kinematically closed channels have been shown to have sizable
influence on the open ones in some other cases~\cite{unitary,ttv}.

The $K$ matrix for the four coupled channels can be written as
\begin{eqnarray}
K  \,=\,  K^{\rm T}  \,=\,
\left( \begin{array}{cccc}   \displaystyle
K_{\rm oo}^{}  &  K_{\rm oc}^{}
\vspace{1ex} \\   \displaystyle
K_{\rm co}^{}  &  K_{\rm cc}^{}
\end{array} \right)   \,\,,
\end{eqnarray}
where the subscripts ``o'' and ``c'' refer to open and closed
channels, respectively, at  $\,\sqrt{s}=m_\Omega^{}.\,$
Thus  $K_{\rm oo,oc,co,cc}$  are all 2$\times$2 matrices in this
case and   $\,K_{\rm co}^{}=K_{\rm oc}^{\rm T}.\,$
Now, it is convenient to introduce the matrix
\begin{eqnarray}
K_{\rm r}^{}  \,=\,
K_{\rm oo}^{} + {\rm i}K_{\rm oc}^{}
\bigl( \openone-{\rm i}q_{\rm c}^{} K_{\rm cc}^{} \bigr)^{-1}
q_{\rm c}^{} K_{\rm co}^{}  \,\,,
\end{eqnarray}
where  $\openone$ is the  2$\times$2 unit matrix  and
$\,q_{\rm c}^{}={\rm diag} \bigl(k_{\Sigma\bar{K}},k_{\Xi\eta}^{}\bigr),\,$
with  $k_{B\phi}^{}$  being the magnitude of the CM three-momentum
in  $B\phi$ scattering, implying that  $k_{\Sigma\bar{K}}$  and
$k_{\Xi\eta}^{}$  are purely imaginary at  $\,\sqrt{s}=m_\Omega^{}.\,$
The elements of  $\cal S$  in  Eq.~(\ref{S})  can then be evaluated
using~\cite{pilkuhn}
\begin{eqnarray}
{\cal S}  \,=\,  \openone + 2{\rm i}\, q_{\rm o}^{1/2}\,
K_{\rm r}^{} \bigl( \openone-{\rm i}q_{\rm o}^{} K_{\rm r}^{} \bigr)^{-1}\,
q_{\rm o}^{1/2}   \,\,,
\end{eqnarray}
where
$\,q_{\rm o}^{} = q_{\rm o}^{1/2} q_{\rm o}^{1/2} =
{\rm diag} \bigl( k_{\Xi\pi}^{}, k_{\Lambda\bar{K}} \bigr) .\,$
For the  $K$-matrix elements, we make the simplest approximation
by adopting the partial-wave amplitudes  $f_{B\phi\to B'\phi'}$
at leading order in chiral perturbation theory, namely
\begin{eqnarray}
\begin{array}{c}   \displaystyle
K_{\rm oo}^{}  \,=\,
\left( \begin{array}{cccc}   \displaystyle
f_{\Xi\pi\to\Xi\pi}^{} & f_{\Xi\pi\to\Lambda\bar{K}}
\vspace{1ex} \\   \displaystyle
f_{\Lambda\bar{K}\to\Xi\pi} & f_{\Lambda\bar{K}\to\Lambda\bar{K}}
\end{array} \right)   \,\,,
\vspace{2ex} \\   \displaystyle
K_{\rm oc}^{}  \,=\,  K_{\rm co}^{\rm T}  \,=\,
\left( \begin{array}{cccc}   \displaystyle
f_{\Xi\pi\to\Sigma\bar{K}} &  f_{\Xi\pi\to\Xi\eta}^{}
\vspace{1ex} \\   \displaystyle
f_{\Lambda\bar{K}\to\Sigma\bar{K}} & f_{\Lambda\bar{K}\to\Xi\eta}
\end{array} \right)   \,\,,
\vspace{2ex} \\   \displaystyle
K_{\rm cc}^{}  \,=\,
\left( \begin{array}{cccc}   \displaystyle
f_{\Sigma\bar{K}\to\Sigma\bar{K}} & f_{\Sigma\bar{K}\to\Xi\eta}
\vspace{1ex} \\   \displaystyle
f_{\Xi\eta\to\Sigma\bar{K}} & f_{\Xi\eta\to\Xi\eta}^{}
\end{array} \right)   \,\,.
\end{array}
\end{eqnarray}
Before deriving them, we remark that  time-reversal invariance of
the strong interaction implies
$\,f_{B\phi\to B'\phi'}=f_{B'\phi'\to B\phi}.\,$

The chiral Lagrangian that describes the interactions of the lowest-lying
mesons and baryons is written down in terms of the lightest meson-octet,
baryon-octet, and baryon-decuplet fields~\cite{bsw,JenMan}.
The meson and baryon octets are collected into  $3\times3$  matrices
$\phi$  and  $B$,  respectively, and the decuplet fields are
represented by the Rarita-Schwinger tensor  $T_{abc}^\mu$,
which is completely symmetric in its SU(3) indices ($a,b,c$).
The octet mesons enter through the exponential
$\,\Sigma=\xi^2=\exp({\rm i}\phi/f),\,$  where  $f$  is the pion-decay
constant.

In the heavy-baryon formalism~\cite{JenMan}, the baryons in
the chiral Lagrangian are described by velocity-dependent fields,
$B_v^{}$  and  $T_v^\mu$.
For the strong interactions, the Lagrangian at lowest order in
the derivative and  $m_s^{}$  expansions is given
by~\cite{JenMan,L2refs}
\begin{eqnarray}   \label{Ls}
{\cal L}_{\rm s}^{}  &=&
\left\langle \bar{B}_v^{}\,
{\rm i}v\cdot{\cal D} B_v^{} \right\rangle
+ 2 D \left\langle \bar{B}_v^{} S_v^\mu
 \left\{ {\cal A}_\mu^{}, B_v^{} \right\} \right\rangle
+ 2 F \left\langle \bar{B}_v^{} S_v^\mu
 \left[ {\cal A}_\mu^{}, B_v^{} \right] \right\rangle
\nonumber \\ && \!\!
-\,\,
\bar{T}_v^\mu\, {\rm i}v\cdot{\cal D} T_{v\mu}^{}
+ \Delta m\, \bar{T}_v^\mu T_{v\mu}^{}
+ {\cal C} \left( \bar{T}_v^\mu {\cal A}_\mu^{} B_v^{}
                 + \bar{B}_v^{} {\cal A}_\mu^{} T_v^\mu \right)
\nonumber \\ && \!\!
+\,\,
\frac{b_D^{}}{2 B_0^{}} \left\langle \bar B_v^{}
\left\{ \chi_+^{}, B_v^{} \right\} \right\rangle
+ \frac{b_F^{}}{2 B_0^{}} \left\langle \bar B_v^{}
 \left[ \chi_+^{}, B_v^{} \right] \right\rangle
+ \frac{b_0^{}}{2 B_0^{}} \left\langle \chi_+^{} \right\rangle
 \left\langle \bar B_v^{} B_v^{} \right\rangle
\nonumber \\ && \!\!
+\,\,
\frac{c}{2 B_0^{}}\, \bar T_v^\mu \chi_+^{} T_{v\mu}^{}
- \frac{c_0^{}}{2 B_0^{}} \left\langle \chi_+^{} \right\rangle
 \bar T_v^\mu T_{v\mu}^{}
\,\,+\,\,
\mbox{$\frac{1}{4}$} f^2 \left\langle \chi_+^{} \right\rangle
\,\,+\,\,  \cdots  \,\,,
\end{eqnarray}
where  $\,\langle\cdots\rangle\,$  denotes   $\,{\rm Tr}(\cdots)\,$  in
flavor-SU(3) space, and we have shown only the relevant terms.
In the first two lines,  $S_v^{}$  is the spin operator and
$\,{\cal A}_\mu^{}=\frac{\rm i}{2}
\left( \xi\, \partial_\mu^{}\xi^\dagger
      - \xi^\dagger\, \partial_\mu^{}\xi \right),\,$
with further details given in Ref.~\cite{atv}.
The last two lines of  ${\cal L}_{\rm s}^{}$ contain
$\,\chi_+^{}=\xi^\dagger\chi\xi^\dagger+\xi\chi^\dagger\xi,\,$  with
$\,\chi=2B_0^{} M=2B_0^{}\,{\rm diag}\bigl(m_u^{},m_d^{},m_s^{}\bigr),\,$
which explicitly breaks chiral symmetry.
We will take the isospin limit  $\,m_u^{}=m_d^{}\equiv\hat{m}\,$  and
consequently
$\,\chi={\rm diag}\bigl(m_\pi^2,m_\pi^2,2 m_K^2-m_\pi^2\bigr).\,$
The constants  $D$, $F$, $\cal C$, $B_0^{}$, $b_{D,F,0}^{}$, $c$, $c_0^{}$
are free parameters which can be fixed from data.

\begin{figure}[hb]
\includegraphics{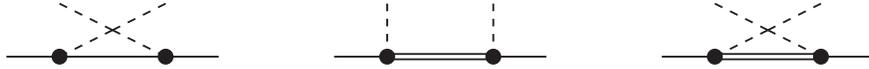}   \vspace{-3ex}
\caption{\label{Pwave}%
Diagrams contributing to the $P$-wave  $\,J=\frac{3}{2}\,$
amplitude for  $\,B\phi\to B'\phi'\,$  at leading order in  $\chi$PT.
In all figures, a dashed line denotes a meson field, a single (double)
solid-line denotes an octet-baryon (decuplet-baryon) field, and each
solid vertex is generated by  ${\cal L}_{\rm s}^{}$  in  Eq.~(\ref{Ls}).}
\end{figure}

In the center-of-mass (CM) frame, the $P$-wave amplitude for
$\,B\phi\to B'\phi'\,$  with total angular-momentum  $J$  has the form
\begin{eqnarray}   \label{M(BphiB'phi')}
{\cal M}_{B\phi\to B'\phi'}^{}  &=&
-8\pi\sqrt{s}\,\, \chi_{B'}^\dagger \left\{
\left[ f_{B\phi\to B'\phi'}^{(P,J=\frac{1}{2})}
      + 2 f_{B\phi\to B'\phi'}^{(P,J=\frac{3}{2})} \right]
\hat{k}{}'\cdot\hat{k}
+
\left[ f_{B\phi\to B'\phi'}^{(P,J=\frac{1}{2})}
      - f_{B\phi\to B'\phi'}^{(P,J=\frac{3}{2})} \right]
{\rm i}\bm{\sigma}\cdot\hat{k}{}'\times\hat{k}
\right\} \chi_{B}^{}   \,\,,
\nonumber \\
\end{eqnarray}
where  $\sqrt{s}$ is the CM energy,  $\chi_B^{}$  and
$\chi_{B'}^{}$  are the Pauli spinors of the baryons,
$\hat{k}$ and  $\hat{k}{}'$  denote the unit vectors
of the momenta of $B$ and $B'$, respectively, and
$f_{B\phi\to B'\phi'}^{(P,J)}$  are the partial-wave amplitudes.
At lowest order in $\chi$PT, the  $\,J=\frac{3}{2}\,$  amplitude
arises from the Lagrangian in Eq.~(\ref{Ls}),  and the pertinent
diagrams are displayed in Fig.~\ref{Pwave}.
The amplitudes in the  $\,I=\frac{1}{2}\,$  channels are then
extracted using the  $\,I=\frac{1}{2}\,$  states in
Eq.~(\ref{|Xp>})  and
\begin{eqnarray}
\begin{array}{c}   \displaystyle
\bigl| \Lambda\bar{K} \bigr\rangle  \,=\,
\bigl| \Lambda K^- \bigr\rangle    \,\,,
\hspace{3em}
\bigl| \Sigma\bar{K} \bigr\rangle  \,=\,
\sqrt{\mbox{$\frac{2}{3}$}}\, \bigl| \Sigma^-\bar{K}{}^0 \bigr\rangle
+ \mbox{$\frac{1}{\sqrt3}$}\, \bigl| \Sigma^0 K^- \bigr\rangle   \,\,,
\hspace{3em}
|\Xi\eta\rangle  \,=\,  \bigl| \Xi^-\eta \bigr\rangle    \,\,,
\end{array}
\end{eqnarray}
which follow a phase convention consistent with the structure
of the  $\phi$  and  $B_v^{}$  matrices.
We write the results as
\begin{eqnarray}   \label{f(P)}
f_{B\phi\to B'\phi'}^{(P,J=\frac{3}{2})}  \,=\,
-{\cal P}_{B\phi,B'\phi'}^{}\,
\frac{k_{B\phi}^{} k_{B'\phi'}^{}\,\sqrt{m_B^{}m_{B'}^{}}}
     {4\pi\,f^2\,\sqrt{s}}   \,\,,
\end{eqnarray}
where  the expressions for  ${\cal P}_{B\phi,B'\phi}^{}$
corresponding to the four channels  have been collected in
Appendix~\ref{PD}.

Since a $D$-wave amplitude has to be at least of second order in
momentum, ${\cal O}\bigl(k^2\bigr)$,  it cannot arise from
the Lagrangian in Eq.~(\ref{Ls}) alone.
Also required is the Lagrangian involving baryons at second order in
the derivative expansion, namely
\begin{eqnarray}   \label{Ls'}
{\cal L}_{\rm s}'  \,=\,
{-1\over 2 m_0^{}}\, \bar{B}_v^{}\,
\bigl[ {\cal D}^2-(v\cdot{\cal D})^2 \bigr] B_v^{}
\,+\,
\frac{1}{2 m_0^{}}\, \bar{T}{}_v^\mu\,
\bigl[{\cal D}^2-(v\cdot{\cal D})^2 \bigr] T_{v\mu}^{}
\,\,+\,\,  \cdots   \,\,,
\end{eqnarray}
where  $m_0^{}$  is the octet-baryon mass in the chiral limit, and
we have shown only the relevant terms.
These are two of the relativistic-correction terms in
the ${\cal O}\bigl(k^2\bigr)$  Lagrangian, and so their coefficients
are fixed.

\begin{figure}[t]
\includegraphics{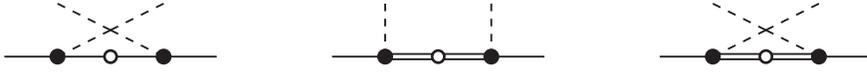}   \vspace{-3ex}
\caption{\label{Dwave}%
Diagrams for the leading nonzero contribution to the $D$-wave
$\,J=\frac{3}{2}\,$  amplitude for  $\,B\phi\to B'\phi'.\,$
Each hollow vertex is generated by  ${\cal L}_{\rm s}'$  in
Eq.~(\ref{Ls'}).}
\end{figure}

In the CM frame, the $D$-wave amplitude for $\,B\phi\to B'\phi'\,$
has the form
\begin{eqnarray}   \label{M'(BphiB'phi')}
{\cal M}_{B\phi\to B'\phi'}'  &=&
-8\pi\sqrt{s}\,\, \chi_{B'}^\dagger \left\{
\left[ 2\, f_{B\phi\to B'\phi'}^{(D,J=\frac{3}{2})}
      + 3\, f_{B\phi\to B'\phi'}^{(D,J=\frac{5}{2})} \right]
\Bigl[ \mbox{$\frac{3}{2}$} \bigl( \hat{k}{}'\cdot
                                  \hat{k}\bigr)^2
      - \mbox{$\frac{1}{2}$} \Bigr]
\right.
\nonumber\\
&& \hspace*{6em}
+ \left.
\left[ f_{B\phi\to B'\phi'}^{(D,J=\frac{3}{2})}
      - f_{B\phi\to B'\phi'}^{(D,J=\frac{5}{2})} \right]
\bigl( 3\hat{k}{}'\cdot\hat{k} \bigr)\,
{\rm i}\bm{\sigma}\cdot\hat{k}{}'\times\hat{k}
\right\} \chi_B^{}   \,\,.
\end{eqnarray}
The leading nonzero contribution to this amplitude for
$\,J=\frac{3}{2}\,$  comes from diagrams shown in Fig.~\ref{Dwave}.
The resulting  $\,I=\frac{1}{2}\,$  partial-wave amplitudes
are given by
\begin{eqnarray}   \label{f(D)}
f_{B\phi\to B'\phi'}^{(D,J=\frac{3}{2})}  \,=\,
-{\cal D}_{B\phi,B'\phi}^{}\,
\frac{k_{B\phi}^2 k_{B'\phi'}^2\, \sqrt{m_B^{}m_{B'}^{}}}{
      4\pi\,f^2\, m_0^{}\, \sqrt{s}}   \;,
\end{eqnarray}
where the expressions for  ${\cal D}_{B\phi,B'\phi}^{}$
corresponding to the four channels have also been collected in
Appendix~\ref{PD}.

Numerically, we adopt the tree-level values
$\,D=0.80\,$  and  $\,F=0.50,\,$  extracted from hyperon semileptonic
decays~\cite{JenMan}, as well as  $\,{\cal C}=-1.7,\,$  from
the strong decays  $\,T\to B\phi.$\footnote{We have chosen the sign
of  $\cal C$  after nonrelativistic quark models~\cite{JenMan}, which
predict  $\,3F=2D\,$  and  $\,{\cal C}=-2D,\,$  both well satisfied
by the adopted  $D$, $F$, and $\cal C$ values.}
We also employ  $\,f=f_\pi^{}=92.4\,{\rm MeV},\,$
$\,m_0^{}=0.7\,\rm GeV,$\footnote{This  $m_0^{}$  value comes
from simultaneously fitting the tree-level formulas for
the octet-baryon masses and the sigma term,
$\,\sigma_{\pi N}^{}=-2 \bigl(b_D^{}+b_F^{}+2 b_0^{}\bigr) \hat{m},\,$
all  derived from Eq.~(\ref{Ls}), to the measured masses and
the empirical value~\cite{sigma}
$\,\sigma_{\pi N}^{}\simeq45\,\rm MeV.\,$
}
and the isospin-averaged masses
\begin{eqnarray}   \label{masses}
\begin{array}{c}   \displaystyle
m_\pi^{}  \,=\,  137.3  \,\,,   \hspace{2em}
m_K^{}  \,=\,  495.7   \,\,,   \hspace{2em}
m_\eta^{}  \,=\,  547.3  \,\,,   \hspace{2em}
\vspace{1ex} \\   \displaystyle
m_N^{}  \,=\,  938.9   \,\,,  \hspace{2em}
m_\Lambda^{}  \,=\,  1115.7   \,\,,  \hspace{2em}
m_\Sigma^{}  \,=\,  1193.2  \,\,,   \hspace{2em}
m_\Xi^{}  \,=\,  1318.1  \,\,,   \hspace{2em}
\vspace{1ex} \\   \displaystyle
m_\Delta^{}  \,=\,  1232.0   \,\,,   \hspace{2em}
m_{\Sigma^*}^{}  \,=\,  1384.6   \,\,,   \hspace{2em}
m_{\Xi^*}^{}  \,=\,  1533.4   \,\,,  \hspace{2em}
m_\Omega^{}  \,=\,  1672.5   \,\,,
\end{array}
\end{eqnarray}
all in units of MeV.
Thus, putting together all the results above and setting
$\,\sqrt{s}=m_\Omega^{},\,$  from the $P$- and $D$-wave
$\cal S$-matrices we obtain
\begin{eqnarray}   \label{deltaPD}
\begin{array}{c}   \displaystyle
\delta_{\Lambda K}^{P} \,=\, -0.65^\circ   \,\,,
\hspace{3em}
\sqrt{\varepsilon_P^{}}   \,=\, 0.013   \,\,,
\hspace{3em}
\delta_{\Lambda K}^{D} \,=\, +0.05^\circ   \,\,,
\hspace{3em}
\sqrt{\varepsilon_D^{}}  \,=\,  0.0009   \,\,,
\end{array}   
\end{eqnarray}
which are pertinent to  Eqs.~(\ref{AO}) and~(\ref{DeltaO}).
The effects of the closed channels turn out to be significant on
$\delta_{\Lambda K}^{P}$  and  $\varepsilon_P^{}$.
Excluding the  $\Sigma\bar{K}$  and  $\Xi\eta$  channels  would
lead to  $\,\delta_{\Lambda K}^{P}=-2.7^\circ\,$  and
$\,\sqrt{\varepsilon_P^{}}=0.065.\,$
The closed channels have minor effects on the $D$-wave parameters.

Since the numbers in Eq.~(\ref{deltaPD}) proceed from the leading
nonzero amplitudes in $\chi$PT, part of the uncertainties in these
predictions comes from our lack of knowledge about the higher-order
contributions, which are presently incalculable.
To get an idea of how they might affect our results, we redo
the calculation using the one-loop values  $\,D=0.61,\,$
$\,F=0.40,\,$  and  $\,{\cal C}=-1.2\,$~\cite{JenMan,bss},
finding  $\,\delta_{\Lambda K}^{P}=-0.47^\circ,\,$
$\,\sqrt{\varepsilon_P^{}}=0.010,\,$
$\,\delta_{\Lambda K}^{D}=+0.03^\circ,\,$  and
$\,\sqrt{\varepsilon_D^{}}=0.0003.\,$
The differences between the two sets of results then provide
an indication of the size of this part of the uncertainties.
Another part is due to our lack of knowledge about the reliability
of our $K$-matrix approximation.
A comparison of $K$-matrix results in $\Lambda\pi$ scattering with
experiment suggests that this  approach gives results with
the correct order-of-magnitude and sign~\cite{ttv,dukes}.
For these reasons, we may conclude that
\begin{eqnarray}   \label{DeltaPD}
-0.9^\circ  \,\le\,  \delta_{\Lambda K}^P-\delta_{\Lambda K}^D
\,\le\,  -0.5^\circ   \,\,,
\hspace{2em}
0.01  \,\le\, \sqrt{\varepsilon_P^{}}  \,\le\, 0.02   \,\,,
\hspace{2em}
0.0003  \,\le\, \sqrt{\varepsilon_D^{}}  \,\le\, 0.002   \,\,.
\hspace*{2em}
\end{eqnarray}
We will employ these numbers in evaluating the asymmetries.

\section{$\bm{CP}$-violating asymmetries within standard model\label{A_sm}}

To calculate  the $CP$-violating phases,  we will work in
the framework of heavy-baryon $\chi$PT.
The amplitude for the weak decay  $\,\Omega\to B\phi\,$
in the heavy-baryon approach has the general form
\begin{eqnarray}   \label{iM}
{\rm i} {\cal M}_{\Omega\to B\phi}^{} \,=\,
-{\rm i} \bigl\langle B\phi \bigr| {\cal L} \bigl| \Omega \bigr\rangle
\;=\;
\bar{u}_{B}^{} \left( \vphantom{|_|^|}
{\cal A}_{B\phi}^{(P)}+2S_v^{}\cdot k_\phi^{}\,{\cal A}_{B\phi}^{(D)}\,
\right) k_\phi^{}\cdot{u}_\Omega^{}   \;.
\end{eqnarray}
where $k_\phi^{}$  is the four-momentum of  $\phi$, and
the superscripts refer to the  $P$- and $D$-wave components of
the amplitude.
In the rest frame of $\Omega$, these components are related to the $p$
and $d$ amplitudes~by
\begin{eqnarray}   \label{p,d}
p  \;=\;  \bigl| \bm{k}_\phi^{} \bigr| \, {\cal A}^{(P)}   \;,
\hspace{3em}
d  \;=\;  \bm{k}_\phi^2\, {\cal A}^{(D)}   \;.
\end{eqnarray}
We will follow the usual prescription for estimating a weak
phase~\cite{dhp,hcpvt',tv4}, namely,  first calculating the imaginary
part of the amplitude and then dividing it by the real part of
the amplitude extracted from experiment under the assumption of no
strong phases and no $CP$ violation.

Within the SM, the weak interactions responsible for hyperon
nonleptonic decays are described by the short-distance effective
$\,|\Delta S|=1\,$  Hamiltonian~\cite{BucBL}
\begin{eqnarray}   \label{Hw_sm}
{\cal H}_{\rm w}^{}  \,=\,
\frac{G_{\rm F}^{}}{\sqrt2}\, V_{ud}^* V_{us}^{}\,
\sum_{i=1}^{10} C_i^{}\, Q_i^{}
\,\,+\,\,  {\rm H.c}   \,\,,
\end{eqnarray}
where  $V_{kl}^{}$ are the elements of
the Cabibbo-Kobayashi-Maskawa (CKM) matrix~\cite{ckm},
\begin{eqnarray}
C_i^{}  \,\equiv\,  z_i^{} + \tau y_i^{}  \,\equiv\,
z_i^{}  \,-\,
\frac{V_{td}^* V_{ts}^{}}{V_{ud}^* V_{us}^{}}\,\, y_i^{}
\end{eqnarray}
are the Wilson coefficients,  and  $Q_i^{}$  are four-quark
operators whose expressions can be found in Ref.~\cite{BucBL}.
In this case, the weak phases  $\phi^{P,D}$  of Eq.~(\ref{AO})
proceed from the $CP$-violating phase residing in the CKM matrix,
and its elements appearing in  $C_i^{}$  above can be expressed in
the Wolfenstein parametrization~\cite{wolfenstein} as
\begin{eqnarray}
V_{ud}^* V_{us}^{}  \,=\,  \lambda   \,\,,
\hspace{3em}
V_{td}^* V_{ts}^{}  \,=\,  -A^2 \lambda^5\, (1-\rho+{\rm i}\eta)
\end{eqnarray}
at lowest order in  $\lambda$.
As is well known,  ${\cal H}_{\rm w}^{}$  transforms mainly as
$\bigl(8_{\rm L}^{},1_{\rm R}^{}\bigr)\oplus
\bigl(27_{\rm L}^{},1_{\rm R}^{}\bigr)$
under  SU(3$)_{\rm L}^{}$$\times$SU(3$)_{\rm R}^{}$  rotations.
It is also known from experiment that the octet term dominates
the 27-plet term~\cite{dgh}.
We, therefore, assume in what follows that within the SM the decays
of interest are  completely characterized by
the  $(8_{\rm L}^{},1_{\rm R}^{})$,  $\,|\Delta I|=\frac{1}{2}\,$
interactions.
The leading-order chiral Lagrangian for such interactions
is~\cite{bsw,jenkins2}
\begin{eqnarray}   \label{Lw_sm}
{\cal L}_{\rm w}^{}  &=&
h_D^{} \left\langle \bar B_v^{} \left\{
\xi^\dagger h \xi\,,\,B_v^{} \right\} \right\rangle
+ h_F^{} \left\langle \bar B_v^{} \left[
\xi^\dagger h \xi\,,\,B_v^{} \right] \right\rangle
+ h_C^{}\, \bar T_v^\mu\, \xi^\dagger h \xi\, T_{v\mu}^{}
\,\,+\,\,  {\rm H.c.}   \,\,,
\end{eqnarray}
where the 3$\times$3-matrix $h$ selects out $\,s\to d\,$ transitions,
having elements  $\,h_{kl}^{}=\delta_{k2}^{}\delta_{3l}^{},\,$  and
the parameters  $h_{D,F,C}^{}$  contain the weak phases of interest.
These phases are induced primarily by the imaginary part of $C_6^{}$
associated with the penguin operator  $Q_6^{}$, and this is due to
its chiral structure and the relative size of  ${\rm Im}\,C_6^{}$.
In order to relate the imaginary part of  $h_{D,F,C}^{}$  to
${\rm Im}\,C_6^{}$,  we use the results of Ref.~\cite{tv4}, obtained
from factorizable and nonfactorizable contributions. 
Accordingly, we have  
\begin{eqnarray}   \label{Imh}
{\rm Im}\, h_D^{}  \,=\,  5.14 \,\, y_6^{}   \,\,,
\hspace{2em}
{\rm Im}\, h_F^{}  \,=\,  -14.3 \,\,  y_6^{}   \,\,,
\hspace{2em}
{\rm Im}\, h_C^{}  \,=\,  32.5 \,\, y_6^{}   \,\,,
\end{eqnarray}  
all in units of
$\,\sqrt2\,f_\pi^{} G_{\rm F}^{} m_{\pi^+}^2\,A^2\lambda^5\eta.\,$

\begin{figure}[b]
\includegraphics{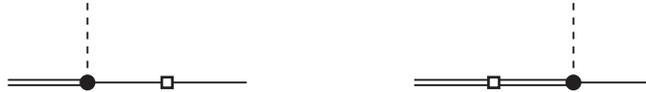}   \vspace{-3ex}
\caption{\label{Pwave_sm}%
Diagrams representing standard-model contributions to the leading-order
$P$-wave amplitude  for  $\,\Omega^-\to B\phi.\,$
Each square represents a weak vertex generated by
${\cal L}_{\rm w}^{}$  in  Eq.~(\ref{Lw_sm}).}
\end{figure}

From  ${\cal L}_{\rm w}^{}$  together with  ${\cal L}_{\rm s}^{}$,
we can derive the diagrams displayed in Fig.~\ref{Pwave_sm},
which represent the leading-order contributions to the $P$-wave
transitions in  $\,\Omega^-\to\Lambda\bar{K},\Xi\pi\,$  and yield
the amplitudes
\begin{eqnarray}   \label{A_P}
\begin{array}{c}   \displaystyle
{\cal A}_{\Lambda\bar{K}}^{(P)}  \,=\,
\frac{{\cal C}\, \bigl( h_D^{}-3h_F^{} \bigr) }
     {2\sqrt{3}\, f\, \bigl( m_\Xi^{}-E_{\Lambda}^{} \bigr) }
- \frac{{\cal C}\, h_C^{}}
       {2\sqrt{3}\,f\,\bigl(m_{\Omega}^{}-m_{\Xi^*}^{}\bigr)}   \,\,,
\vspace{2ex} \\   \displaystyle
{\cal A}_{\Xi\pi}^{(P)}  \,=\,  \sqrt{\mbox{$\frac{2}{3}$}}\, {\cal
A}_{\Xi^0\pi^-}^{(P)} + \mbox{$\frac{1}{\sqrt3}$}\, 
{\cal A}_{\Xi^-\pi^0}^{(P)}
\,=\, \frac{-{\cal C}\, h_C^{}}
     {2\sqrt3\, f\, \bigl( m_{\Omega}^{}-m_{\Xi^*}^{} \bigr) }   \,\,.
\end{array}
\end{eqnarray}
Applying  Eq.~(\ref{Imh})  in
$\,p_{B\phi}^{}=\bigl|\bm{k}_\phi^{}\bigr|\,{\cal A}_{B\phi}^{(P)}\,$
then leads to
\begin{eqnarray}   \label{Imp_sm/px}
\frac{{\rm Im}\,p_{\Lambda\bar{K}}^{}}{p_\Lambda^{\rm expt}}  \,=\,
-1.15\,\, A^2 \lambda^5\eta\,\, y_6^{}   \,\,,
\hspace{3em}
\frac{{\rm Im}\,p_{\Xi\pi}^{}}{p_\Xi^{\rm expt}}  \,=\,
+23.6\,\, A^2 \lambda^5\eta\,\, y_6^{}   \,\,,
\end{eqnarray}
where  $p_{\Lambda,\Xi}^{\rm expt}$  are the central values
of  $p_{\Lambda,\Xi}^{}$ in  Eq.~(\ref{px,dx}).
The uncertainties in these predictions are due to our neglect of
higher-order terms that are presently incalculable and to
our lack of knowledge on the reliability of the matrix-element
calculation.
Therefore, we assign an error of 100$\%$ to these ratios,
as was similarly done in Ref.~\cite{tv4} for the weak phases in 
$\,\Lambda\to p\pi\,$  and  $\,\Xi\to\Lambda\pi.\,$  
Thus, using  $\,A^2 \lambda^5\eta = 1.26\times10^{-4}\,$  and
$\,y_6^{} = -0.096,\,$  as in  Ref.~\cite{tv4}, we obtain
\begin{eqnarray}   \label{phiP_sm}
\phi_\Lambda^P  \,=\,  (1.4\pm 1.4)\times 10^{-5}   \,\,,
\hspace{3em}
\phi_\Xi^P  \,=\,  (-2.9\pm 2.9)\times 10^{-4}   \,\,.
\end{eqnarray}  
The  $\phi_\Xi^P$  result is comparable in size to that
estimated in  Ref.~\cite{tv2} using the vacuum-saturation
method.\footnote{The numerical differences between the estimates also
arise from the use of a positive $\cal C$ value in Ref.~\cite{tv2}.}

Turning now to the $D$-wave phases, we note that the expression for
the  ${\cal A}^{(D)}$  term in Eq.~(\ref{iM}) implies that
${\cal L}_{\rm w}^{}$,  in conjunction with  ${\cal L}_{\rm s}^{}$
and  ${\cal L}_{\rm s}'$,  cannot solely give rise to diagrams for
the $D$-wave components.
Rather, the weak Lagrangian that can generate the leading nonzero
contributions to this term must have the Dirac structure
$\,\bar{B}{}_v^{}S_v^\mu\, \partial_\mu^{}{\cal A}_\alpha^{}\,
T_v^\alpha,\,$
which is of  ${\cal O}\bigl(k^2\bigr)$.
The $D$-wave amplitude at  ${\cal O}\bigl(k^2\bigr)$  can also
receive contributions from so-called tadpole diagrams, each
being a combination of a strong $\,\Omega B\phi\bar{K}$
vertex, generated by a Lagrangian having the structure
$\,\bar{B}{}_v^{}S_v^\mu{\cal A}_\mu^{}{\cal A}_\alpha^{}T_v^\alpha,\,$
and a $\bar{K}$-vacuum  vertex coming from
a weak Lagrangian of  ${\cal O}\bigl(m_s^{}\bigr)$.
Unfortunately, at present the parameters of these strong and weak
Lagrangians of  ${\cal O}\bigl(k^2\bigr)$ are incalculable.
The best that we can do is to make a crude estimate based on  naive
dimensional analysis~\cite{nda}.
Thus, since  the lowest-order chiral Lagrangian yielding  $p_{B\phi}^{}$
is of  ${\cal O}(1)$,  whereas that yielding  $d_{B\phi}^{}$  is of
${\cal O}\bigl(k^2\bigr)$,  and since  $\,k\sim m_s^{}\,$  in hyperon
nonleptonic decays,  we expect that
\begin{eqnarray}   \label{d/p}
\frac{d_{B\phi}^{}}{p_{B\phi}^{}}  \,\sim\,
\frac{m_s^2}{\Lambda_\chi^2}   \,\,,
\end{eqnarray}
where  $\,\Lambda_\chi\sim4\pi f\,$  is the chiral-symmetry
breaking scale.
It is worth remarking here that for  
$\,m_s^{}\sim 0.12\,\rm GeV\,$~\cite{buras2} this naive expectation
is compatible with the value of
$d_\Lambda^{}/p_\Lambda^{}$  from Eq.~(\ref{px,dx}), in which
the  $d_\Lambda^{}$  number is determined largely by the preliminary
data from HyperCP~\cite{lu}.
For these reasons, we make the approximation
\begin{eqnarray}
\phi^D  \,=\, \frac{{\rm Im}\,d}{d^{\rm expt}}  \,=\,
\frac{m_s^2}{\Lambda_\chi^2}\,
\frac{p^{\rm expt}}{d^{\rm expt}}\, \phi^P
\end{eqnarray}
for the magnitude of the phase, where $\phi^P$ comes from
Eq.~(\ref{phiP_sm}).
Since  $d_\Xi^{}$  as quoted in Eq.~(\ref{px,dx}) is poorly determined,
we take the further approximation
$\,d_\Xi^{}=p_\Xi^{} d_\Lambda^{}/p_\Lambda\,$  for its magnitude
in order to estimate  $\phi_\Xi^D$.
All this leads to
\begin{eqnarray}   \label{phiD_sm}
\phi_\Lambda^D  \,=\,  (0\pm3)\times 10^{-5}   \,\,,
\hspace{3em}   
\phi_\Xi^D  \,=\,  (0\pm6)\times 10^{-4}   \,\,.
\end{eqnarray}
The errors that we quote in  $\phi_B^{P,D}$  are obviously not Gaussian
and simply indicate the ranges resulting from our calculation.

Putting together the numbers from  Eqs.~(\ref{px,dx}), (\ref{DeltaPD}),
(\ref{phiP_sm}), and~(\ref{phiD_sm})  in  Eq.~(\ref{AO})  yields   
\begin{eqnarray}
-9\times10^{-6}  \,\le\,  A_\Omega^{}  \,\le\,
+2\times10^{-6}   \,\,.
\end{eqnarray}
We note that the second term on the right-hand side of  Eq.~(\ref{AO}),  
which would vanish if the  $\,\Xi\pi\leftrightarrow\Lambda\bar{K}\,$  
rescattering were ignored, has turned out to be the largest one.  
This is due to  $\phi_\Xi^P$  and  $\varepsilon_P^{}$  being much 
larger than  $\phi_\Lambda^{P,D}$  and  $\varepsilon_D^{}$, 
respectively, as well as to  $\delta_{\Lambda K}^{P,D}$  being small.  
For the partial-rate asymmetry in Eq.~(\ref{DeltaO}), we find
\begin{eqnarray}   \label{Delta_O^sm}
0  \,\le\,  \Delta_\Omega^{}  \,\le\,  13\times10^{-6}   \,\,.
\end{eqnarray}   
This is comparable to the corresponding asymmetry in
$\,\Omega\to\Xi\pi\,$~\cite{tv2}, but larger than those in
octet-hyperon decays~\cite{dhp}.

Since the asymmetry measured by HyperCP is the sum
$\,A_{\Omega\Lambda}^{}=A_\Omega^{}+A_\Lambda^{},\,$  it is important
to know how  $A_\Omega^{}$  compares with  $A_\Lambda^{}$.
The SM contribution to  $A_\Lambda^{}$  has been evaluated
most recently to be
$\,-3\times10^{-5}\le A_\Lambda^{}\le 4\times10^{-5}\,$~\cite{tv4}.   
Thus within the standard model  $A_\Omega^{}$  is smaller than
$A_\Lambda^{}$, but not negligibly so, and the resulting
$A_{\Omega\Lambda}^{}$  has a value within the range
\begin{eqnarray}   \label{A_OL^sm}
-4\times10^{-5}  \,\le\,  A_{\Omega\Lambda}^{}  \,\le\,  
4\times10^{-5}   \,\,.
\end{eqnarray}
For this observable, HyperCP expects to have a statistical
precision of  $\,9\times10^{-2}\,$~\cite{lu},  and so its
measurement will unlikely be sensitive to the SM effects.

\section{$\bm{CP}$-violating asymmetries due to new physics\label{A_np}}

Here we evaluate  $A_\Omega^{}$  and  $\Delta_\Omega^{}$  arising
from possible physics beyond the standard model.
In particular, we consider contributions generated by
the chromomagnetic-penguin operators (CMO), which in some
new-physics models could be significantly larger that their SM
counterparts~\cite{NP,sdg,buras3}.
The relevant effective Hamiltonian can be written as~\cite{buras3}
\begin{eqnarray}   \label{Hw_np}
{\cal H}_{{\rm w},g}^{}  \,=\,
C_{g}^{}\, Q_g^{}  \,+\,  \tilde{C}_{g}^{}\, \tilde{Q}{}_g^{}
\,\,+\,\,  {\rm H.c.}   \,\,,
\end{eqnarray}
where $C_{g}^{}$  and  $\tilde{C}_{g}^{}$  are the Wilson
coefficients, and
\begin{eqnarray}
Q_{g}^{}  \,=\,
\frac{g_{\rm s}^{}}{16\pi^2}\, \bar{d}\, \sigma^{\mu\nu} t^a\,
\bigl(1+\gamma_5^{}\bigr) s\, G_{\!\mu\nu}^{a}   \,\,,
\hspace{3em}
\tilde{Q}{}_{g}^{}  \,=\,
\frac{g_{\rm s}^{}}{16\pi^2}\, \bar{d}\, \sigma^{\mu\nu} t^a\,
\bigl(1-\gamma_5^{}\bigr) s\, G_{\!\mu\nu}^{a}
\end{eqnarray}
are the CMO, with  $G_a^{\!\mu\nu}$  being the gluon field-strength
tensor,  $g_{\rm s}^{}$  the gluon coupling constant,  and
$\,{\rm Tr}\bigl(t^a t^b\bigr)=\frac{1}{2}\delta^{ab}.\,$
Since various new-physics scenarios may contribute differently to
the coefficients of the operators, we will not focus on specific models,
but will instead adopt a model-independent approach, only assuming that
the contributions are potentially sizable, in order to estimate bounds
on the resulting asymmetries as allowed by constraints from kaon
measurements.

The chiral Lagrangian proceeding from the CMO has to respect their
symmetry properties.  
Under  $\,\rm SU(3)_{L}^{}$$\times$$\rm SU(3)_{R}^{}\,$  rotations
$Q_g^{}$ and $\tilde{Q}{}_g^{}$  transform as
$\,\bigl(\bar{3}{}_{\rm L}^{},3_{\rm R}^{}\bigr)\,$  and
$\,\bigl(3_{\rm L}^{},\bar{3}{}_{\rm R}^{}\bigr),\,$  respectively.
Moreover, under a $CPS$ transformation  (a $CP$  operation
followed by interchanging the~$s$ and~$d$ quarks)
$Q_g^{}$  and  $\tilde{Q}{}_g^{}$  change into each other.
These symmetry properties are also those of the quark densities
$\,\bar{d}(1\pm\gamma_5^{})s,\,$  of which the lowest-order
chiral realization has been derived in Ref.~\cite{tv4}.
From this realization, we can infer the leading-order chiral
Lagrangian induced by the CMO, namely
\begin{eqnarray}   \label{Lw_np}
{\cal L}_{{\rm w},g}^{}  &=&
\beta_D^{} \left\langle \bar{B}{}_v^{}
\left\{ \xi^\dagger h\xi^\dagger, B_v^{} \right\} \right\rangle
+ \beta_F^{} \left\langle \bar{B}{}_v^{}
 \left[ \xi^\dagger h\xi^\dagger, B_v^{} \right] \right\rangle
+ \beta_0^{} \left\langle h\Sigma^\dagger \right\rangle
\left\langle \bar{B}{}_v^{} B_v^{} \right\rangle
\nonumber \\ && \!\!
+\,\,
\tilde\beta_D^{} \left\langle \bar{B}{}_v^{}
\left\{ \xi h\xi,B_v^{} \right\} \right\rangle
+ \tilde\beta_F^{} \left\langle \bar{B}{}_v^{}
 \left[ \xi h\xi,B_v^{} \right] \right\rangle
+ \tilde\beta_0^{} \left\langle h\Sigma \right\rangle
 \left\langle \bar{B}{}_v^{} B_v^{} \right\rangle
\nonumber \\ && \!\!
+\,\,
\beta_C^{}\, \bar{T}{}_v^\alpha\, \xi^\dagger h\xi^\dagger\,
T_{v\alpha}^{}
- \beta_0' \left\langle h\Sigma^\dagger \right\rangle
 \bar{T}{}_v^\alpha T_{v\alpha}^{}
+ \tilde\beta_C^{}\, \bar{T}{}_v^\alpha\, \xi h\xi\, T_{v\alpha}^{}
- \tilde\beta_0'
 \left\langle h\Sigma \right\rangle \bar{T}{}_v^\alpha T_{v\alpha}^{}
\nonumber \\ && \!\!
+\,\,
\beta_\varphi^{}\, f^2 B_0^{}
\left\langle h\Sigma^\dagger \right\rangle
\,\,+\,\,
\tilde\beta_\varphi^{}\, f^2 B_0^{}
\left\langle h\Sigma \right\rangle
\,\,+\,\,  {\rm H.c.}    \,\,,
\end{eqnarray}
where  $\beta_i^{}$  $\bigl(\tilde\beta_i^{}\bigr)$  are parameters
containing the coefficient  $C_g^{}$  $\bigl(\tilde{C}_g^{}\bigr)$.
The part of this Lagrangian without the decuplet-baryon fields was
first written down in Ref.~\cite{t6}.

\begin{figure}[t]
\includegraphics{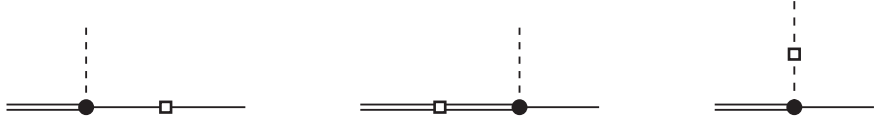}   \vspace{-3ex}
\caption{\label{Pwave_np}%
Diagrams representing chromomagnetic-penguin contributions to
the leading-order $P$-wave amplitude for  $\,\Omega^-\to B\phi.\,$
Each square represents a weak vertex generated by
${\cal L}_{\rm w}^{\rm CMO}$  in  Eq.~(\ref{Lw_np}).}
\end{figure}

From  ${\cal L}_{{\rm w},g}$  along with  ${\cal L}_{\rm s}$,
we derive the diagrams shown in Fig.~\ref{Pwave_np}, which
represent the lowest-order contributions induced by the CMO to
the $P$-wave transitions in  $\,\Omega\to\Lambda\bar{K},\Xi\pi.\,$
We remark that each of the three diagrams in the figure is
of~${\cal O}(1)$  in the  $m_s^{}$  expansion, and
that Fig.~\ref{Pwave_sm} does not include the meson-pole diagram
because within the SM it contributes only at next-to-leading order.
The amplitudes following from Fig.~\ref{Pwave_np} are
\begin{eqnarray}   \label{AP_np}
\begin{array}{c}   \displaystyle
{\cal A}_{\Lambda\bar{K}}^{(P)g}  \,=\,
\frac{{\cal C}\, \bigl( \beta_D^{+}-3\beta_F^{+} \bigr) }
     {2\sqrt{3}\, f\, \bigl( m_\Xi^{}-E_\Lambda^{} \bigr) }
- \frac{{\cal C}\, \beta_C^{+}}
       {2\sqrt{3}\, f\, \bigl(m_{\Omega}^{}-m_{\Xi^*}^{}\bigr)}   \,\,,
\vspace{2ex} \\   \displaystyle
{\cal A}_{\Xi\pi}^{(P)g}  \,=\,
\frac{-{\cal C}\, \beta_C^{+}}
     {2\sqrt3\, f\, \bigl( m_{\Omega}^{}-m_{\Xi^*}^{} \bigr) }
+ \frac{\sqrt3\, {\cal C}\, \beta_\varphi^+}
       {2f\, \bigl(m_s^{}-\hat{m}\bigr)}    \,\,,
\end{array}
\end{eqnarray}
where  $\,\beta_i^{+}\equiv\beta_i^{}+\tilde{\beta}_i^{}\,$  and
we have used  $\,m_K^2-m_\pi^2=B_0^{}\, (m_s^{}-\hat m),\,$  derived
from Eq.~(\ref{Ls'}).\footnote{
It is worth noting here that, as in the  $\,\Lambda\to p\pi\,$  and
$\,\Xi\to\Lambda\pi\,$  cases~\cite{t6},  each of the two amplitudes
in  Eq.~(\ref{AP_np})  vanishes if we set
$\,\beta_{D,F}^{+}=\kappa^{+} b_{D,F}^{},\,$
$\,\beta_C^{+}=\kappa^{+} c,\,$  and
$\,\beta_\varphi^{+}=\kappa^{+}/2,\,$  with  $\kappa^+$  being a constant,
take the limit  $\,E_\Lambda^{}=m_\Lambda^{},\,$  and use the relations
$\,m_\Xi^{}-m_\Lambda^{}=\frac{2}{3} \bigl( b_D^{}-3 b_F^{} \bigr)
\bigl( m_s^{}-\hat m \bigr) \,$
and
$\,m_\Omega^{}-m_{\Xi^*}^{}=\frac{2}{3}\,c\,\bigl( m_s^{}-\hat m \bigr),\,$
both derived from Eq.~(\ref{Ls}).
This satisfies the requirement implied by the Feinberg-Kabir-Weinberg
theorem~\cite{fkw} that the operator  $\,\bar{d}s\,$  cannot contribute
to physical decay amplitudes~\cite{DonGH1}, and thus serves
as a check for the formulas in Eq.~(\ref{AP_np}).
}

In order to estimate the weak phases in $A_\Omega^{}$, we
need to determine the parameters  $\beta_i^{+}$  in terms of
the underlying coefficient  $\,C_g^{+}\equiv C_g^{}+\tilde{C}_g^{},\,$
which is the combination corresponding to parity-conserving transitions.
From the effective Hamiltonian in Eq.~(\ref{Hw_np}) and the chiral
Lagrangian in Eq.~(\ref{Lw_np}), we can derive the one-particle matrix
elements
\begin{eqnarray}   \label{<B'|H|B>}
\begin{array}{c}   \displaystyle
\bigl\langle n\bigr|{\cal H}_{{\rm w},g}^{}\bigl|\Lambda\bigr\rangle
\,=\,
\frac{\beta_D^{+}+3\beta_F^{+}}{\sqrt6}\, \bar{u}_n^{}u_\Lambda^{}  \,\,,
\hspace{3em}
\bigl\langle\Lambda\bigr|{\cal H}_{{\rm w},g}^{}\bigl|\Xi^0\bigr\rangle
\,=\,
\frac{\beta_D^{+}-3\beta_F^{+}}{\sqrt6}\, \bar{u}_\Lambda^{}u_\Xi^{}   \,\,,
\vspace{2ex} \\   \displaystyle
\bigl\langle\Xi^{*-}\bigr|{\cal H}_{{\rm w},g}^{}\bigl|\Omega^-\bigr\rangle
\,=\,
\frac{-\beta_C^{+}}{\sqrt3}\, \bar{u}{}_{\Xi^*}^{}\cdot u_\Omega^{}  \,\,,
\hspace{3em}
\bigl\langle\pi^-\bigr|{\cal H}_{{\rm w},g}^{}\bigl|K^-\bigr\rangle
\,=\,  \beta_\varphi^{+}\, B_0^{}   \,\,.
\end{array}
\end{eqnarray}
Since there is presently no reliable way to determine these matrix
elements from first principles, we employ the MIT bag model to
estimate them.   
The results for  $\beta_{D,F,\varphi}^{+}$  have already been 
derived in  Ref.~\cite{t6}  using the bag-model calculations of 
Ref.~\cite{dghp} and are given by    
\begin{eqnarray}   \label{bdbf}
\begin{array}{c}   \displaystyle
\beta_D^+  \,=\,  -\mbox{$\frac{3}{7}$}\, \beta_F^+  \,=\,
\frac{2\, I_M^{}N^4}{\pi\, R^2}\, C_g^+   \,\,,
\hspace{3em}
\beta_\varphi^{+}  \,=\,
\frac{-8\, I_M^{}N^4\, \sqrt{2m_K^2}}{\pi B_0^{}\, R^2}\, C_g^+   \,\,,
\end{array}
\end{eqnarray}
where  $N$, $R$, and $I_M^{}$  are bag parameters.
For  $\beta_C^+$,  extending the work of Ref.~\cite{dghp} we find
\begin{eqnarray}   \label{bc}
\beta_C^+  \,=\,  \frac{-8\, I_M^{}N^4}{\pi\, R^2}\, C_g^+   \,\,.
\end{eqnarray}
Numerically,  we take  $\,R=5.0\,{\rm GeV}^{-1}\,$  for the octet baryons,
$\,R=5.4\,{\rm GeV}^{-1}\,$  for the decuplet baryons,  and
$\,R=3.3\,{\rm GeV}^{-1}\,$  for the mesons, after
Refs.~\cite{dghp,bagmodel}.
In addition, as in Ref.~\cite{t6}, we have  $\,N=2.27\,$  and
$\,I_M^{}=1.63\times10^{-3}\,$   for both the baryons and mesons.
It follows that
\begin{eqnarray}   \label{beta_i}
\begin{array}{c}   \displaystyle
\beta_D^+  \,=\,  -\mbox{$\frac{3}{7}$} \beta_F^+  \,=\,
1.10\times10^{-3}\,\, C_g^+\,\,{\rm GeV}^2   \,\,,
\hspace{3em}  
\beta_C^+  \,=\,  -3.78\times10^{-3}\,\, C_g^+\,\,{\rm GeV}^2   \,\,,
\vspace{2ex} \\   \displaystyle
\beta_\varphi^+\, B_0^{}  \,=\,
-7.09\times10^{-3}\,\, C_g^+\,\, {\rm GeV}^3   \,\,,
\end{array}
\end{eqnarray}
We note that  $C_g^{+}$  here is the Wilson coefficient at the low scale
$\,\mu={\cal O}(1\,\rm GeV)\,$  and hence  already contains
the QCD running from the new-physics scales.
We also note that the bag-model numbers in Eq.~(\ref{beta_i}) are
comparable in magnitude to the natural values of the parameters as
obtained from naive dimensional analysis~\cite{nda},
\begin{eqnarray}   \label{beta_i^nda}
\beta_{D,F,C}^{\rm NDA}  \,=\,
\frac{C_g^{}\, g_{\rm s}^{}}{16\pi^2}\, \frac{\Lambda_\chi^2}{4\pi}
\,\sim\,  0.0024\,C_g^{}\,\, {\rm GeV}^2   \,\,,
\hspace{3em}  
\beta_\varphi^{\rm NDA}\, B_0^{}  \,=\,  
{C_g^{}\, g_{\rm s}^{}\over 16\pi^2}\, \frac{\Lambda_\chi^3}{4\pi}  
\,\sim\,  0.0028\,C_g^{}\,\, {\rm GeV}^3   \,\,,
\end{eqnarray}
where   we have chosen  $\,g_{\rm s}^{}=\sqrt{4\pi}.\,$
The differences between the two sets of numbers provide an indication
of the level of uncertainty in estimating the matrix elements..
This will be taken into account in our results below.

Applying  Eq.~(\ref{beta_i})  in
$\,p_{B\phi}^{}=\bigl|\bm{k}_\phi^{}\bigr|\,{\cal A}_{B\phi}^{(P)}\,$
then leads to the CMO contributions 
\begin{eqnarray}   \label{phiP_np}
\bigl( \phi_\Lambda^P \bigr)_g  \,=\,
(-1.0\pm2.0)\times10^5\,{\rm GeV}\,\, {\rm Im}\, C_g^+   \,\,,
\hspace{3em}
\bigl( \phi_\Xi^P \bigr)_g  \,=\,
(2.3\pm4.6)\times10^5\,{\rm GeV}\,\, {\rm Im}\, C_g^+     \,\,.
\end{eqnarray}
where, as in the  $\,\Lambda\to p\pi\,$  and  $\,\Xi\to\Lambda\pi\,$  
cases~\cite{t6},  we have assigned an error of 200$\%$ to each of these
numbers to reflect the uncertainty due to our neglect of
higher-order terms that are presently incalculable and
the uncertainty in estimating the matrix elements above.
For the $D$-wave phases, we have here the same problem in estimating 
them as in the standard-model case, and so we have to
resort again to dimensional arguments.
Thus, since the $D$-wave amplitude is parity violating, we have
\begin{eqnarray}   \label{phiD_np}
\bigl( \phi_\Lambda^D \bigr)_g  \,=\,
(0\pm 3)\times10^5\,{\rm GeV}\,\, {\rm Im}\, C_g^-   \,\,,
\hspace{3em}  
\bigl( \phi_\Xi^D \bigr)_g  \,=\,
(0\pm 8)\times10^5\,{\rm GeV}\,\, {\rm Im}\, C_g^-   \,\,,  
\end{eqnarray}
where  $\,C_g^{-}\equiv C_g^{}-\tilde{C}_g^{}\,$  is
the combination corresponding to parity-violating transitions.

Putting together the numbers from  Eqs.~(\ref{px,dx}), (\ref{DeltaPD}),
(\ref{phiP_np}), and~(\ref{phiD_np})  in  Eq.~(\ref{AO}), we find
\begin{eqnarray}
10^{-4}\, {\rm GeV}^{-1}\, \bigl( A_\Omega^{} \bigr)_g  \,=\,
(0.3\pm 1.3)\, {\rm Im}\, C_g^+  +  (0\pm 1)\, {\rm Im}\, C_g^-   \,\,.
\end{eqnarray}  
As in the SM result, the second term in $A_\Omega^{}$ dominates these
numbers.  
For the partial-rate asymmetry, we obtain  
\begin{eqnarray}   \label{Delta_O^np}
\bigl( \Delta_\Omega^{} \bigr)_g^{}  \,=\,
(-0.7\pm 1.4)\times10^4\, {\rm GeV}\,\, {\rm Im}\, C_g^+   \,\,.
\end{eqnarray}
We can now write down the contribution of the CMO to the sum of asymmetries
$\,A_{\Omega\Lambda}^{}=A_\Omega^{}+A_\Lambda^{}\,$  being measured by
HyperCP.  
The most recent evaluation of their contribution to  $A_\Lambda^{}$  has
been done in Ref.~\cite{t6}, the result being
$\,10^{-4}\, {\rm GeV}^{-1}\, (A_\Lambda^{})_g^{} =  
(-4.2\pm 8.3)\, {\rm Im}\,C_g^{+} + (3.5\pm 7.0)\, {\rm Im}\,C_g^{-}.\,$  
Evidently,  $\bigl(A_\Omega^{}\bigr){}_g$  is much smaller than, 
though still not negligible compared to,  $\bigl(A_\Lambda^{}\bigr){}_g$.  
Summing the two asymmetries yields
\begin{eqnarray}   \label{A_OL^np}
10^{-4}\, {\rm GeV}^{-1}\, \bigl(A_{\Omega\Lambda}^{}\bigr)_g  \,=\,
(-4\pm 10)\, {\rm Im}\, C_g^{+} + (4\pm 8)\, {\rm Im}\,C_g^{-}   \,\,.
\end{eqnarray}

Since the CMO also contribute to the $CP$-violating parameters
$\epsilon$ in kaon mixing and $\epsilon'$ in kaon decay, which are
now well measured, it is possible to obtain bounds on
$\bigl(A_{\Omega\Lambda}^{}\bigr){}_g^{}$  and
$\bigl(\Delta_\Omega^{}\bigr){}_g^{}$
using the  $\epsilon$  and  $\epsilon'$  data.
As discussed in Ref.~\cite{t6}, the experimental values
$\,|\epsilon|=(22.80\pm 0.13)\times10^{-4}\,$   and
$\,{\rm Re}(\epsilon'/\epsilon)=(16.6\pm1.6)\times10^{-4}\,$~\cite{buras2,pdb}
imply that
\begin{eqnarray}
\bigl|{\rm Im}C_g^+\bigr|  \,<\,  5.0\times10^{-8}\,{\rm GeV}^{-1}   \,\,,
\hspace{3em}
\bigl|{\rm Im}C_g^-\bigr|  \,<\,  7.4 \times10^{-9}\,{\rm GeV}^{-1}   \,\,.
\end{eqnarray}
Then, from Eqs.~(\ref{Delta_O^np}) and~(\ref{A_OL^np}), it follows that  
\begin{eqnarray}   \label{(AOL,DeltaO)g}
\bigl|A_{\Omega\Lambda}^{}\bigr|_g  \,<\,  8\times10^{-3}   \,\,,
\hspace{3em}
\bigl|\Delta_\Omega^{}\bigr|_g  \,<\,  1\times10^{-3}   \,\,.
\end{eqnarray}
The upper limits of these ranges well exceed those within
the SM in Eqs.~(\ref{Delta_O^sm})  and~(\ref{A_OL^sm}), but
the largest size of  $\bigl(A_{\Omega\Lambda}^{}\bigr){}_g^{}$
is still an order of magnitude below the expected sensitivity of
HyperCP~\cite{lu}.  
This, nevertheless, implies that a nonzero measurement by HyperCP would 
be an unmistakable signal of new physics.

\section{Conclusion\label{conclusion}}
    
We have evaluated the sum of the $CP$-violating asymmetries
$A_\Omega^{}$  and  $A_\Lambda^{}$  occurring in the decay chain
$\,\Omega\to\Lambda K\to p\pi K,\,$  which is currently being
studied by the HyperCP experiment.
The dominant contribution to  $A_\Omega^{}$  has turned out to be due
to final-state interactions via  $\,\Omega\to\Xi\pi\to\Lambda K.\,$
We have found that both within and beyond the standard model
$A_\Omega^{}$  is smaller than  $A_\Lambda^{}$, but not negligibly so.
Taking a model-independent approach, we have also found that 
contributions to  $\,A_{\Omega\Lambda}^{}=A_\Omega^{}+A_\Lambda^{}$
from possible new-physics through the chromomagnetic-penguin operators 
are allowed by constraints from kaon data to exceed the SM effects by 
up to two orders of magnitude.
In summary, 
\begin{eqnarray*}  
\bigl| A_{\Omega\Lambda}^{} \bigr|_{\rm SM}  \,\le\,
4\times10^{-5}   \,\,, 
\hspace{3em}  
\bigl| A_{\Omega\Lambda}^{} \bigr|_g  \,<\,  8\times10^{-3}    \,\,.  
\end{eqnarray*}   
Since the SM contribution is well beyond the expected reach of HyperCP, 
a finding of nonzero asymmetry would definitely indicate 
the presence of new physics.  
In any case, the upcoming data on  $A_{\Omega\Lambda}^{}$  will yield 
information which complements that to be gained from the measurement of 
$A_{\Xi\Lambda}^{}$  in  $\,\Xi\to\Lambda\pi\to p\pi\pi.\,$

Finally, we have shown that the contribution of
$\,\Omega\to\Xi\pi\to\Lambda K\,$  also causes the partial-rate
asymmetry  $\Delta_\Omega^{}$  in  $\,\Omega\to\Lambda K\,$  to be
nonvanishing, thereby providing another means to observe $CP$  
violation in this decay.  
This asymmetry and that in  $\,\Omega\to\Xi\pi\,$  tend to be
larger than the corresponding asymmetries in octet-hyperon decays 
and hence are potentially useful probes of $CP$ violation  
in future experiments.  
Since  $\Delta_\Omega^{}$  results from the interference of
$P$-wave amplitudes, a~measurement of it will probe  
the underlying parity-conserving interactions.  
Numerically, we have found   
\begin{eqnarray*}   
0  \,\le\,  \bigl( \Delta_\Omega^{} \bigr)_{\rm SM}  \,\le\,  
1\times10^{-5}   \,\,,  
\hspace{3em}   
\bigl| \Delta_\Omega^{} \bigr|_g  \,<\,  1\times10^{-3}   \,\,.
\end{eqnarray*}
where the bound on the contribution of the CMO arises from
the constraint imposed by  $\epsilon$  data.

\begin{acknowledgments}    
I would like to thank G.~Valencia for helpful discussions and comments.
I am also grateful to E.C.~Dukes and L.-C.~Lu for experimental information.
This work was supported in part by the Lightner-Sams Foundation.
\end{acknowledgments}

\appendix

\section{$\bm{{\cal P}}$  and $\bm{{\cal D}}$  factors in
$\bm{P}$-wave  and  $\bm{D}$-wave  $\,\bm{J=\frac{3}{2}}\,$
amplitudes for  $\,\bm{B\phi\to B'\phi'}\,$  in
$\,\bm{I=\frac{1}{2}}$,  $\bm{S=-2}\,$  channels  \label{PD}}

For the four coupled channels, the  ${\cal P}$  factors are
\begin{eqnarray}
\begin{array}{c}   \displaystyle
{\cal P}_{\Xi\pi,\Xi\pi}^{}  \,=\,
\frac{-\frac{1}{6}(D-F)^2}{E_\Xi^{}-E_\pi^\prime-m_\Xi^{}}
+ \frac{\frac{1}{12}\, {\cal C}^2}{\sqrt{s}-m_{\Xi^*}^{}}
+ \frac{-\frac{1}{108}\, {\cal C}^2}{
        E_\Xi^{}-E_\pi^\prime-m_{\Xi^*}^{}}   \,\,,
\vspace{2ex} \\   \displaystyle
{\cal P}_{\Xi\pi,\Lambda\bar{K}}  \,=\,
\frac{\frac{1}{3}D(D+F)}{E_\Xi^{}-E_K^\prime-m_\Sigma^{}}
+ \frac{\frac{1}{12}\, {\cal C}^2}{\sqrt{s}-m_{\Xi^*}^{}}
+ \frac{\frac{1}{36}\, {\cal C}^2}
       {E_\Xi^{}-E_K^\prime-m_{\Sigma^*}^{}}   \,\,,
\vspace{2ex} \\   \displaystyle
{\cal P}_{\Xi\pi,\Sigma\bar{K}}  \,=\,
\frac{-\frac{1}{9}D(D-3 F)}{E_\Xi^{}-E_K^\prime-m_\Lambda^{}}
+ \frac{-\frac{2}{3}(D+F)F}{E_\Xi^{}-E_K^\prime-m_\Sigma^{}}
+ \frac{-\frac{1}{12}\, {\cal C}^2}{\sqrt{s}-m_{\Xi^*}^{}}
+ \frac{\frac{1}{54}\, {\cal C}^2}
       {E_\Xi^{}-E_K^\prime-m_{\Sigma^*}^{}}   \,\,,
\vspace{2ex} \\   \displaystyle
{\cal P}_{\Xi\pi,\Xi\eta}^{}  \,=\,
\frac{-\frac{1}{6}(D-F)(D+3 F)}{E_\Xi^{}-E_\eta^\prime-m_\Xi^{}}
+ \frac{-\frac{1}{12}\, {\cal C}^2}{\sqrt{s}-m_{\Xi^*}^{}}
+ \frac{-\frac{1}{36}\, {\cal C}^2}{
        E_\Xi^{}-E_\eta^\prime-m_{\Xi^*}^{}}   \,\,,
\end{array}
\end{eqnarray}
\begin{eqnarray}
\begin{array}{c}   \displaystyle
{\cal P}_{\Lambda\bar{K},\Lambda\bar{K}}  \,=\,
\frac{\frac{1}{18}(D+3 F)^2}{E_\Lambda^{}-E_K^\prime-m_N^{}}
+ \frac{\frac{1}{12}\, {\cal C}^2}{\sqrt{s}-m_{\Xi^*}^{}}   \,\,,
\vspace{2ex} \\   \displaystyle
{\cal P}_{\Lambda\bar{K},\Sigma\bar{K}}  \,=\,
\frac{-\frac{1}{6}(D-F)(D+3 F)}{E_\Lambda^{}-E_K^\prime-m_N^{}}
+ \frac{-\frac{1}{12}\, {\cal C}^2}{\sqrt{s}-m_{\Xi^*}^{}}   \,\,,
\vspace{2ex} \\   \displaystyle
{\cal P}_{\Lambda\bar{K},\Xi\eta}  \,=\,
\frac{\frac{1}{9}D(D-3 F)}{E_\Lambda^{}-E_\eta^\prime-m_\Lambda^{}}
+ \frac{-\frac{1}{12}\, {\cal C}^2}{\sqrt{s}-m_{\Xi^*}^{}}   \,\,,
\end{array}
\end{eqnarray}
\begin{eqnarray}
\begin{array}{c}   \displaystyle
{\cal P}_{\Sigma\bar{K},\Sigma\bar{K}}  \,=\,
\frac{-\frac{1}{6}(D-F)^2}{E_\Sigma^{}-E_K^\prime-m_N^{}}
+ \frac{\frac{1}{12}\, {\cal C}^2}{\sqrt{s}-m_{\Xi^*}^{}}
+ \frac{\frac{2}{27}\, {\cal C}^2}
       {E_\Sigma^{}-E_K^\prime-m_\Delta^{}}   \,\,,
\vspace{2ex} \\   \displaystyle
{\cal P}_{\Sigma\bar{K},\Xi\eta}  \,=\,
\frac{\frac{1}{3}D(D+F)}{E_\Sigma^{}-E_\eta^\prime-m_\Sigma^{}}
+ \frac{\frac{1}{12}\, {\cal C}^2}{\sqrt{s}-m_{\Xi^*}^{}}
+ \frac{-\frac{1}{36}\, {\cal C}^2}{
        E_\Sigma^{}-E_\eta^\prime-m_{\Sigma^*}^{}}  \,\,,
\end{array}
\end{eqnarray}
\begin{eqnarray}
\begin{array}{c}   \displaystyle
{\cal P}_{\Xi\eta,\Xi\eta}^{}  \,=\,
\frac{\frac{1}{18}(D+3 F)^2}{E_\Xi^{}-E_\eta^\prime-m_\Xi^{}}
+ \frac{\frac{1}{12}\, {\cal C}^2}{\sqrt{s}-m_{\Xi^*}^{}}
+ \frac{\frac{1}{36}\, {\cal C}^2}
       {E_\Xi^{}-E_\eta^\prime-m_{\Xi^*}^{}}   \;,
\end{array}
\end{eqnarray}
and the  ${\cal D}$  factors
\begin{eqnarray}
\begin{array}{c}   \displaystyle
{\cal D}_{\Xi\pi,\Xi\pi}^{}  \,=\,
\frac{(D-F)^2}{60 \bigl( E_\Xi^{}-E_\pi^\prime-m_\Xi^{} \bigr)^2}
- \frac{7\, {\cal C}^2}{
        540 \bigl( E_\Xi^{}-E_\pi^\prime-m_{\Xi^*}^{} \bigr)^2}   \,\,,
\vspace{2ex} \\   \displaystyle
{\cal D}_{\Xi\pi,\Lambda\bar{K}}  \,=\,
\frac{-D(D+F)}{30 \bigl( E_\Xi^{}-E_K^\prime-m_\Sigma^{} \bigr)^2}
+ \frac{7\, {\cal C}^2}{
        180 \bigl( E_\Xi^{}-E_K^\prime-m_{\Sigma^*}^{} \bigr)^2}   \,\,,
\vspace{2ex} \\   \displaystyle
{\cal D}_{\Xi\pi,\Sigma\bar{K}}  \,=\,
\frac{D(D-3 F)}{90 \bigl( E_\Xi^{}-E_K^\prime-m_\Lambda^{} \bigr)^2}
+ \frac{(D+F)F}{15 \bigl( E_\Xi^{}-E_K^\prime-m_\Sigma^{} \bigr)^2}
+ \frac{7\, {\cal C}^2}{
        270 \bigl( E_\Xi^{}-E_K^\prime-m_{\Sigma^*}^{} \bigr)^2}   \,\,,
\vspace{2ex} \\   \displaystyle
{\cal D}_{\Xi\pi,\Xi\eta}^{}  \,=\,
\frac{(D-F)(D+3 F)}{60 \bigl( E_\Xi^{}-E_\eta^\prime-m_\Xi^{} \bigr)^2}
+ \frac{-7\, {\cal C}^2}{
        180 \bigl( E_\Xi^{}-E_\eta^\prime-m_{\Xi^*}^{} \bigr)^2}   \,\,,
\end{array}
\end{eqnarray}
\begin{eqnarray}
\begin{array}{c}   \displaystyle
{\cal D}_{\Lambda\bar{K},\Lambda\bar{K}}  \,=\,
\frac{-(D+3 F)^2}{180 \bigl(E_\Lambda^{}-E_K^\prime-m_N^{}\bigr)^2}   \,\,,
\hspace{3em}
{\cal D}_{\Lambda\bar{K},\Sigma\bar{K}}  \,=\,
\frac{(D-F)(D+3 F)}{60 \bigl( E_\Lambda^{}-E_K^\prime-m_N^{} \bigr)^2}  \,\,,
\vspace{2ex} \\   \displaystyle
{\cal D}_{\Lambda\bar{K},\Xi\eta}  \,=\,
\frac{-D(D-3 F)}{
      90 \bigl( E_\Lambda^{}-E_\eta^\prime-m_\Lambda^{} \bigr)^2}   \,\,,
\end{array}
\end{eqnarray}
\begin{eqnarray}
\begin{array}{c}   \displaystyle
{\cal D}_{\Sigma\bar{K},\Sigma\bar{K}}  \,=\,
\frac{(D-F)^2}{60 \bigl( E_\Sigma^{}-E_K^\prime-m_N^{} \bigr)^2}
+ \frac{14\, {\cal C}^2}{
        135 \bigl( E_\Sigma^{}-E_K^\prime-m_\Delta^{} \bigr)^2}   \,\,,
\vspace{2ex} \\   \displaystyle
{\cal D}_{\Sigma\bar{K},\Xi\eta}  \,=\,
\frac{-D(D+F)}{30 \bigl( E_\Sigma^{}-E_\eta^\prime-m_\Sigma^{} \bigr)^2}
+ \frac{-7\, {\cal C}^2}{
        180 \bigl(E_\Sigma^{}-E_\eta^\prime-m_{\Sigma^*}^{}\bigr)^2}  \,\,,
\end{array}
\end{eqnarray}
\begin{eqnarray}
{\cal D}_{\Xi\eta,\Xi\eta}^{}  \,=\,
\frac{-(D+3 F)^2}{180 \bigl( E_\Xi^{}-E_\eta^\prime-m_\Xi^{} \bigr)^2}
+ \frac{7\, {\cal C}^2}{
        180 \bigl( E_\Xi^{}-E_\eta^\prime-m_{\Xi^*}^{} \bigr)^2}   \;,
\end{eqnarray}
where  $E_\phi'$  is the energy of  $\phi$  in the final state.
We note that contributions to the propagators from the $\Delta m$
and quark-mass terms in Eq.~(\ref{Ls})  have been implicitly included
in these results.

\end{document}